\documentclass[conference]{IEEEtran}
\IEEEoverridecommandlockouts
\usepackage{cite}
\usepackage{amsmath,amssymb,amsfonts}
\usepackage{algorithmic}
\usepackage{graphicx}
\usepackage{textcomp}
\usepackage{xcolor}
\usepackage{paralist}
\usepackage{enumitem}
\usepackage{hyperref}
\usepackage{cleveref}
\usepackage{float}
\usepackage[textsize=footnotesize]{todonotes}
\presetkeys{todonotes}{fancyline}{}
\usepackage{balance}
\Crefname{section}{\S$\!$}{\S\S$\!$}

\def\BibTeX{{\rm B\kern-.05em{\sc i\kern-.025em b}\kern-.08em
    T\kern-.1667em\lower.7ex\hbox{E}\kern-.125emX}}
\begin{document}


\title{Exploring the Environmental Benefits of In-Process Isolation for Software Resilience}

\author{\IEEEauthorblockN{Merve G\"{u}lmez}
\IEEEauthorblockA{\textit{Ericsson Security Research} - Kista, Sweden \\
\textit{imec-Distrinet, KU Leuven} - Leuven, Belgium \\
merve.gulmez@kuleuven.be}
\and
\IEEEauthorblockN{Jan Tobias M\"{u}hlberg}
\IEEEauthorblockA{\textit{Universit\'e Libre de Bruxelles, Ecole Polytechnique} \\
\textit{Cybersecurity Research Center} - Brussels, Belgium \\
jan.tobias.muehlberg@ulb.be}
}

\author{\IEEEauthorblockN{Merve G\"{u}lmez}
\IEEEauthorblockA{\textit{Ericsson Security Research} \\ Kista, Sweden \\
\textit{imec-DistriNet, KU Leuven} \\ Leuven, Belgium \\
merve.gulmez@kuleuven.be}
\and
\IEEEauthorblockN{Thomas Nyman}
\IEEEauthorblockA{\textit{Ericsson Product Security} \\ Jorvas, Finland \\
thomas.nyman\\@ericsson.com}
\and
\IEEEauthorblockN{Christoph Baumann}
\IEEEauthorblockA{\textit{Ericsson Security Research} \\
Kista, Sweden \\
christoph.baumann\\@ericsson.com}
\and
\IEEEauthorblockN{Jan Tobias M\"{u}hlberg}
\IEEEauthorblockA{%
\textit{imec-DistriNet, KU Leuven} \\ Leuven, Belgium \\
\textit{Universit\'e Libre de Bruxelles} \\ Brussels, Belgium \\
jan.tobias.muehlberg@ulb.be}
}

\maketitle

\begin{abstract}
Memory-related errors remain an important cause of software vulnerabilities. 
While mitigation techniques such as using memory-safe languages are promising solutions,
these do not address software resilience and availability. In this paper, we propose a solution 
to build resilience against memory attacks into software,
which contributes to environmental sustainability and security. 
\end{abstract}
\section{Introduction}\label{sec:introduction}
The urgent need for climate action requires a focus on sustainability in every field, 
including security~\cite{DanielGruss}.  
Recent work attempts to define numerous different notions for sustainable
security~\cite{muehlberg_2022_sustaining_sec, Paverd2019SustainableS}. 
For example, Paverd et al.~\cite{Paverd2019SustainableS} define design principles for sustainable security and safety, such as isolation between components, replication,
and diversification. However, these principles relate to sustaining the
security of the system rather than environmental sustainability. In this
paper, we discuss how building resilience into software against memory
attacks can, by limiting the need for redundancy in dependable systems
software, contribute to environmental sustainability in addition to sustaining security. 

Memory-unsafe programming languages such as C, and C++ are still one of the primary root causes of software vulnerabilities~\cite{project0day}. 
These vulnerabilities can allow attackers to access vulnerable systems by compromising program behavior. 
There are two ways to address memory-related attacks: \begin{inparaenum}[1)] \item retrofit the unsafe software with run-time defense techniques, 
\item perform new development in memory-safe languages.\end{inparaenum}

While well-known mitigations such as control-flow integrity and stack canaries
can detect attacks, they stop the attacks by terminating the application. Service-oriented applications are particularly at risk because a temporary failure can have an impact on a large number of clients.
Consequently,  designing systems for resilience additionally requires replication or redundancy.
Such over-provisioning, however, is not environmentally friendly. There is a need to address software resilience through
approaches that are environmentally sustainable and also
built upon principles of sustainable security. 

Rust is an emerging memory-safe language with performance
nearly as good as conventional system programming languages such as C and
C++~\cite{Pereira_2017}. Rust is more difficult to exploit by design, as it forbids unsafe memory access patterns and enforces a strong type system.
However, Rust provides foreign function interfaces (FFI) that enable an
application to use system libraries written in other languages, essentially calling memory-unsafe
code. This can violate the memory safety of Rust code,
e.g., by dereferencing raw pointers or breaking ownership rules~\cite{CLA}.
Even if such applications are mainly written in Rust, there is no guarantee that the application is resilient against memory corruption attacks. 
Efficient approaches for software resilience could benefit Rust applications that rely on unsafe code through FFI as well. 
In earlier research we introduce \emph{Secure Rewind and Discard of Isolated Domains}~\cite{gulmez2022dsn}, as briefly described in~\Cref{sec:sdrad}. 
Our approach leverages mechanisms for hardware-assisted in-process isolation in commercial, off-the-shelf processors and allows retrofitting
C applications with the capability to recover from memory errors in isolated components. 
However, a drawback of this approach is compartmentalizing an application into distinct, 
isolated domains as a pre-requisite for secure rewind can require code changes and manual effort by developers.
That drives up the cost of software development, both in terms of money and energy consumption.

To address this drawback, and to extend the approach to protect
FFI functionality, preserving the memory-safety guarantees of Rust
applications, we are currently adapting our secure rewind library to
a streamlined version that allows developers to leverage
metaprogramming in Rust to annotate functions that must be
compartmentalized, as discussed in~\Cref{sec:sdrad_ffi}.

In future work, we plan to quantify the benefit of secure rewind in
terms of benefits to environmental sustainability gained through
improved resilience against attacks. We discuss possible
methodologies in~\Cref{sec:sustainability}.

\section{Secure Domain Rewind and Discard}\label{sec:sdrad}
Our \emph{Secure Rewind and Discard of Isolated Domains} 
scheme~\cite{gulmez2022dsn} allows restoring the execution state of a 
program
to a state in which allocated memory is free from corruption. This is possible with two pre-requirements: 
\begin{inparaenum}[1)]
    \item Compartmentalizing an application into distinct domains by leveraging hardware-assisted
    software fault isolation which guarantees that a memory defect within a domain only affects that domain's memory.  
    \item Leveraging different pre-existing detection mechanisms, such as stack canaries and domain violations. 
\end{inparaenum}

We realize this idea by providing a C library, \emph{Secure Domain Rewind and Discard
(SDRaD)} for Linux on commodity 64-bit x86 processors with Protection Keys for Userspace (PKU). 
The library provides flexible APIs to support different compartmentalization schemes, such as protecting application integrity and confidentially.
The developer can retrofit software applications with the secure rewinding mechanism by using them.

We evaluated our solution in three different use cases: Memcached, NGINX, and OpenSSL and concluded that it adds negligible overhead (2\%--4\%) 
in realistic multi-processing scenarios.
Our solution can provide faster recovery time compared to container or process restart time. For example, in our Memcached setup with a 10GB database, 
a regular restart takes about 2 minutes, in-process rewinding takes only $3.5\mathit{\mu{}s}$.
Also, our approach offers significant advantages with 
limiting the impact of malicious clients on other clients in a service-oriented application, without disrupting service.

One of the drawbacks of our solution is that retrofitting an application
with SDRaD requires additional software development effort.
Depending on the target application, this effort might be minimal. For example, we changed two source files in Memcached and added 484 new lines
of wrapper code.

\section{Friend or Foe? Foreign Functions in Rust}\label{sec:sdrad_ffi}
One of the shortcomings in Rust's foreign function interface is that it allows
calling unsafe code inside Rust. In an earlier approach, unsafe code was invoked as a separate process with significant run-time overheads~\cite{Lamowski2017}.
Also, several researchers propose heap isolation between the unsafe and safe
blocks based on PKU by using compiler-based approaches~\cite{Kirth22, liu2020}.
However, while both approaches can provide integrity in the memory safe area,
they do not address the availability of software applications.

In ongoing work, we propose a solution, \emph{SDRaD for Foreign Function
Interfaces (SDRaD-FFI)}, to improve the availability and
memory-safety guarantees of software written in Rust without excluding the
use of legacy code. This is
possible by protecting the integrity of the safe memory area from the unsafe
area and allowing the safe area to perform alternate actions in case of memory
violation. This requires lightweight in-process isolation between all safe and
unsafe memory areas.

We will provide a Rust crate as a realization of our approach by leveraging our
SDRaD C library. Moreover, we aim to provide easy-to-use annotations by
leveraging Rust's macro expansion to hide SDRaD calls, argument and return value
handling, and alternate actions in case of domain violations.  SDRaD-FFI can
support arbitrary argument passing between domains using different Rust
serialization crates.  We plan to evaluate different serialization crates and
our solution in real-world use cases.

\section{Sustainability Evaluation}\label{sec:sustainability}
Providing well-defined components and isolation is important for sustaining
security~\cite{Paverd2019SustainableS}. For example, isolation can limit memory corruption within a component. 
However, some approaches for compartmentalizing applications can be less environmentally sustainable than others.
For example, conventional process isolation has high context-switching costs that increase resource utilization.
Hardware-assisted in-process isolation, such as Memory
Protection Keys (MPK)~\cite{Park2023,Park19} and CHERI~\cite{Watson15}, are potential solutions to provide lightweight isolation. 
It should be noted that CHERI requires specialized hardware. 
Implementing these solutions may require refactoring applications, which comes with software development costs. 
Another important issue is that while compartmentalizing applications into distinct domains is important for protecting application integrity and confidentially, providing availability is also important for environmental sustainability. 
Our solution is based on hardware-assisted in-process isolation by leveraging MPK; it is lightweight and incurs minimal run-time overhead. 
SDRaD APIs allow the developer to compartmentalize applications into distinct domains to provide confidentially and integrity. 
Also, the rewinding mechanism is important for application availability. 
While implementing SDRaD requires software development effort, we aim to reduce the effort for SDRaD-FFI with easy-to-use APIs.

Replication or diversification of software can decrease the likelihood of memory-related attacks and increase software longevity. 
This can result in over-provisioning hardware resources and is not environmentally friendly. 
Our solution supports fast recovery time without replication or diversification, and with only limited runtime overhead. 
In fact, SDRaD substantially reduces the time to recover from a fault. For example, in our Memcached setup, 
a regular restart takes about 2 minutes (which would violate 99.999\% availability if there were three faults per year), 
while our in-process rewinding takes only $3.5\mathit{\mu{}s}$, allowing
for more than $9\cdot10^7$ recoveries. Specifically in critical application
scenarios, e.g., in telecommunications or smart grids, high levels of
availability are normally achieved by means of redundancy, which our
approach can alleviate. A thorough analysis of the potential impacts of our
approach requires further life-cycle assessment
approaches with a focus on environmental sustainability through energy
efficiency~\cite{kern2018sustainable,gregoriou2019energy}, but also economic
and social
dimensions~\cite{condori2020action}, to be applied in a comprehensive case
study from the above
domains, which would also consider rebound
effects~\cite{gossart2015rebound}. Our results already show that in
particular systems that involve software components such as 
Memcached, which need a lot of time to recover from failure, can benefit
from SDRaD in terms of dependability and life-cycle sustainability.

\section{Conclusion}
\label{sec:conclusion}\label{sec:future_work}\label{sec:availability}
\label{futurework}
We outline early research on applications and impact assessment of our
\emph{Secure Rewind and Discard of Isolated Domains} approach to improve
software resilience and availability.
We highlight ongoing work to improve on dependability
properties of software written in memory-safe languages that rely on
foreign/native library code. Importantly, our approach has the potential to
also improve on aspects of environmental sustainability in critical software
systems.

\section{Acknowledgments} 
This research is partially funded by the Research Fund KU Leuven, the
Flemish Research Programme Cybersecurity, and the CyberExcellence
programme of the Walloon Region, Belgium.
This research has received funding under EU H2020 MSCA-ITN action 5GhOSTS, 
grant agreement no. 814035.

\balance
\bibliographystyle{plain}
\bibliography{main}

\end{document}